\documentclass[conference]{IEEEtran}
\IEEEoverridecommandlockouts
\usepackage[T1, T2A]{fontenc}
\usepackage{amsmath,amssymb,amsfonts}
\usepackage{algorithmic}
\usepackage{graphicx}
\usepackage{textcomp}
\usepackage{xcolor}
\usepackage{cases}
\usepackage{makecell}
\usepackage{hyperref}
\usepackage{multirow}
\usepackage{substitutefont}
\usepackage{flushend}
\usepackage[backend=biber, style=ieee]{biblatex}
\def\BibTeX{{\rm B\kern-.05em{\sc i\kern-.025em b}\kern-.08em
    T\kern-.1667em\lower.7ex\hbox{E}\kern-.125emX}}
\AtBeginBibliography{\small}

\substitutefont{T2A}{\rmdefault}{Tempora-TLF}
\renewcommand{\rmdefault}{cmr}

\font\tenipa=tipa10
\def\schwa{{\tenipa\char64}}

\begin{document}

\title{
Macedonian Speech Synthesis for \\
Assistive Technology Applications

\thanks{The work was carried out within the project ``Inclusion of children with disabilities in pre-primary education'' funded by the UNICEF office in RN Macedonia.}
}

\author{\IEEEauthorblockN{
    Bojan Sofronievski$^1$, Elena Velovska$^1$, Martin Velichkovski$^1$, Violeta Argirova$^1$, Tea Veljkovikj$^1$, \\
    Risto Chavdarov$^1$, Stefan Janev$^2$, Veselinka Labroska$^3$, Kristijan Lazarev$^4$, Toni Bachvarovski$^4$, \\
    Zoran Ivanovski$^1$, Dimitar Tashkovski$^1$,  and Branislav Gerazov$^1$
}
\vspace{5pt}
\IEEEauthorblockA{
    $^1$\textit{FEEIT, Ss Cyril and Methodius University in Skopje, RN Macedonia} \\
    $^2$\textit{Melon Inc., Skopje, RN Macedonia} \\
    $^3$\textit{Institute of Macedonian language ``Krste Misirkov'' - Skopje, RN Macedonia} \\
    $^4$\textit{Association for Assistive Technologies ``Open the Windows'', Skopje, RN Macedonia} \\
}
\vspace{10pt}
    \small{bojan.sof@ieee.org, elenavelovska@gmail.com, gerazov@feit.ukim.edu.mk}
}

\maketitle
\begin{abstract}
Speech technology is becoming ever more ubiquitous with the advance of speech enabled devices and services.
The use of speech synthesis in Augmentative and Alternative Communication tools, has facilitated inclusion of individuals with speech impediments allowing them to communicate with their surroundings using speech.
Although there are numerous speech synthesis systems for the most spoken world languages, there is still a limited offer for smaller languages.
We propose and compare three models built using parametric and deep learning techniques for Macedonian trained on a newly recorded corpus.
We target low-resource edge deployment for Augmentative and Alternative Communication and assistive technologies, such as communication boards and screen readers.
The listening test results show that parametric speech synthesis is as performant compared to the more advanced deep learning models.
Since it also requires less resources, and offers full speech rate and pitch control, it is the preferred choice for building a Macedonian TTS system for this application scenario.
\end{abstract}


\section{Introduction}

Speech synthesis is the artificial production of human speech. 
It's main use is the generation of intelligible and natural sounding speech on the basis of textual input, i.e. text-to-speech (TTS) synthesis \cite{dutoit1997introduction}. 
Up until the 1990’s, TTS synthesizers predominately used formant synthesis and were based on modelling the speech production process \cite{pinto1989formant}.
Although scoring high on intelligibility, these systems did not sound very natural and came with a high development cost.
This limited the development of TTS technology to only the larger languages with ample resources.
In the 1990's, concatenative synthesis became increasingly popular \cite{
hunt1996unit, 
}.
Based on the concatenation of prerecorded natural speech segments to generate the requested speech output, this approach gives the synthetic speech a very natural sound, whilst eliminating the expensive speech model development process.
This led to the proliferation of TTS systems across the globe, bringing speech synthesis for the first time to many of the languages of the world.
In fact, due to its high naturalness, concatenative synthesis was the predominant technology used in most commercial systems up until the late 2010's \cite{capes2017siri}, and some still use it today.

Concatenative systems, although highly natural, are inflexible in that they allow limited control of pitch and rhythm, if any.
This precluded their use in utility TTS systems, such as screen readers, where control was more important than naturalness.
Also, they require the recording and annotation of large corpora to add new voices.
At the turn of the century, these limitations as well as the strive to unite both TTS and automatic speech recognition (ASR) under one common framework opened the doors to the paradigm of statistical parametric speech synthesis \cite{yoshimura1999simultaneous}.
Parametric TTS systems introduced in the 2000's used the well established Hidden Markov Models (HMMs) and training methods used in ASR to reverse the flow and use them to generate speech based on input text.
This allowed increased system control and eased the creation of new voices through reducing training data requirements. 
Due to its reduced naturalness, parametric synthesis was primarily used in research and assistive technologies.

Since the late 2010's advances in speech synthesis can be overwhelmingly contributed to the development of deep learning models. 
Indeed, today's state-of-the-art TTS systems are almost exclusively based on the superior performance of general purpose end-to-end mappings that use millions of tunable parameters \cite{oord2016wavenet, wang2017tacotron}.
Deep Learning (DL) has led TTS to the long term goal of reaching human naturalness of synthetic speech, so much so that new systems can hardly be identified as artificial \cite{shen2018natural}.

Speech enabled assistive technology is the key enabler of the inclusion of people with disabilities \cite{teixeira2009speech}.
Speech synthesis allows the digital inclusion of the blind or visually impaired through its use in screen readers.
Speech enabled Augmentative and Alternative Communication (AAC) systems allow the speech and language impaired to communicate with their surroundings.
Communication boards, a type of AAC, allow the user to select a sequence of symbols represented via icons and then synthesise speech based on the sequence.
Communication boards can be deployed on mobile electronic devices, such as tablets and smartphones, as well as dedicated hardware.
Cboard is a free software AAC web app that allows users to communicate in a number of languages\footnote{\url{https://www.cboard.io}}.
The speech synthesis systems presented in this paper were developed within a larger project for localising Cboard for Macedonian.
Previously, Govorko – a communication board targeting Macedonian users was developed by students at FEEIT \cite{govorko}.\footnote{\url{https://gitlab.com/govorko/govorko}}

The first attempts at speech synthesis in Macedonian date back to 1996, when a concept solution for such a system was proposed.
Early attempts include emulating Macedonian using the Croatian diphone inventory under the MBROLA framework, \cite{grcevskiSSMK, zrmanovskaSSMK}.
The first Macedonian TTS systems developed used diphone based concatenative synthesis \cite{gerazov08diphone, chungurski2009govorni}.
At that time Macedonian was also included in regional commercial offerings, initially using a Serbian inventory \cite{delic2006review}, and later a dedicated one \cite{delic2010speech}.
The recent availability of free and open source implementations of Deep Learning based speech synthesis systems has stimulated the development of both commercial\footnote{\url{https://maika.mk/}} and open source \cite{mishev2020makedonka} TTS systems for Macedonian in 2020.
Our results from 2021, confirm that DL models can indeed approach human naturalness in speech quality.\footnote{\url{https://speech.feit.ukim.edu.mk/}}
In mid 2021 we saw the release of the first fully functional free software based TTS for Macedonian, with its inclusion in the RHVoice parametric speech synthesis system.\footnote{\url{https://rhvoice.org/}}

We train and compare four speech synthesis models targeting low-resource edge deployment for use in assistive speech technology, primarily Augmentative and Alternative Communication applications and screen readers.
To train the models we design and record a speech corpus with a Macedonian professional speaker.
We evaluate the models via an online listening test using the Mean Opinion Score (MOS) and Multiple Stimuli with Hidden Reference and Anchor (MUSHRA) method.

\section{Text normalisation for standard Macedonian}

Macedonian is a South Slavic language spoken by more than 3 million native speakers in Macedonia and in the world.
Grapheme-to-phoneme mapping and stress position are an important part of the text normalisation module that enables high-quality speech synthesis in Macedonian.
Macedonian orthography is essentially phonemic, eliminating the need for complex grapheme-to-phoneme mapping.
Each of the 31 phonemes is represented by a single grapheme.
In addition there are 6 allophones for 4 of the phonemes, most notably syllabic r, which comprises almost exclusively all the occurrences of the schwa /\schwa/.
These allophones are determined by the context of the occurrence of the grapheme.
Another exception to the one-to-one grapheme-to-phone mapping in Macedonian are the two rules of orthoepy: voicing assimilation and devoicing \cite{friedman2002macedonian}.
Voicing assimilation occurs when two consonants with different voicing come next to each other in a word.
In this case, the first consonant is substituted for its voiced/voiceless pair, in suit to the voicing of the second consonant.
Although this process is largely integrated into Macedonian orthography, there are still exceptions.
Devoicing, on the other hand, is never written and occurs on the ultimate voiced consonant at the end of phrase boundaries. 

Stress in Macedonian is largely antepenultimate -- occurring on the 3rd syllable from the last.
Thus the absolute position of the stress shifts if the word is extended with suffixes.
For words with 2 syllables this translates to the first syllable being stressed.
Exceptions to this rule are almost exclusively made up of loan words, but also include some Macedonian words and personal names.
Not all loan words have irregular stress though, as some have either been accommodated to the antepenultimate rule, while some are in the process of accommodation and can bear both an irregular and regular stress. 

\section{Methodology}

In this work, we train and evaluate one parametric speech synthesis system and three Deep Learning systems based on one TTS and two neural vocoder models.
For the parametric TTS we use RHVoice, while we base the DL systems on the Tacotron 2 \cite{shen2018natural} TTS model to generate the mel spectrograms from the input text.
We then use three different vocoders to generate the speech waveform:
\begin{itemize}
    \item the original Griffin-Lim algorithm \cite{griffin1984signal} for phase estimation, followed by an inverse short-time Fourier transform, as in the original Tacotron approach,
    \item the Multi-band MelGAN vocoder \cite{mbmelgan}, and
    \item the Parallel WaveGAN vocoder \cite{pwgan}.
\end{itemize}
We chose to use these vocoders based on our target real-time edge deployment scenario.
There are higher quality vocoders, such as the autoregressive WaveNET \cite{oord2016wavenet}, however they cannot run in real-time, especially not on devices with limited resources.

To measure and compare the quality of the synthesised speech with the different models, we conducted a subjective Mean Opinion Score (MOS) test as well as a comparative Multiple Stimuli Hidden Reference and Anchor (MUSHRA).

\subsection{RHVoice and HTS}

RHVoice is an open-source multilingual speech synthesizer focused on bringing free high-quality voices to the visually impaired for use with their screen reader. 
Their focus is on languages for which there is a lack of good voices for this purpose.
The creator of RHVoice, Olga Yakovleva, and many of the contributors are blind or partially sighted themselves.
The speech synthesis engine is based on the HMM speech synthesis system HTS \cite{zen2007hmm}, which in turn uses the HMM toolkit HTK used primarily for speech recognition \cite{young2002htk}.
The system uses context-dependent HMMs to simultaneously model the speech spectrum, the excitation signal, and phoneme durations. 
The target features are the Mel Generalised Cepstrum and band aperiodicities, and the system also models the pitch via the log $f_0$.
Although the synthesised speech lacks human level naturalness, the synthesis is fully controllable, allowing for changes in speech rate and pitch, thus making it perfect for the use with screen readers.

\subsection{Tacotron 2}

Tacotron 2 has been one of the most successful Deep Learning models developed for for speech synthesis \cite{shen2018natural}.
It is the system to achieve human-level naturalness, synthesising speech that is difficult to distinguish from real human speech.
The system comprises a sequence-to-sequence architecture based on the original Tacotron \cite{wang2017tacotron}, modified to generate mel spectrograms, and a WaveNet based neural vocoder \cite{oord2016wavenet}.
The Tacotron model simplifies the traditional speech synthesis pipeline by eliminating the need of complex linguistic and acoustic features with a single neural network trained from data alone. 

\subsection{Multi-band MelGAN}


Generative adversarial networks (GANs) have shown great success in many computer vision related tasks, such as image generation and image-to-image translation, but neural vocoders like MelGAN \cite{melgan}, Parallel WaveGAN \cite{pwgan} and multi-band MelGAN \cite{mbmelgan}, have shown very promising performance on audio waveform generation tasks.
These models provide very fast waveform generation, making them very suitable for real-time applications.
MelGAN is a non-autoregressive feed-forward convolutional neural network architecture that perform audio waveform generation \cite{melgan}.
It was the first successfully trained GAN to yield high-quality TTS synthesis in real-time on a CPU. 

The generator in MelGAN uses a stack of transposed convolutional layers to upsample the input Mel sequence. 
Each transposed convolution is followed by a stack of residual blocks with dilated convolutions to increase the receptive field.
The success of MelGAN is essentially achieved by using multiple discriminators at different audio scales, motivated from the fact that audio has structure at different levels and each discriminator intends to learn features for different frequency ranges of audio.
In addition to the discriminator’s signal, a feature matching objective is also used to train the generator.
This objective minimizes the L1 distance between the discriminator feature maps of real and synthetic audio.

Multi-band MelGAN (MB-MelGAN) is an architecture evolved from the basic MelGAN, providing even faster waveform generation and quality improvements in the generated speech \cite{mbmelgan}.
The generator network in MB-MelGAN generates signals in multiple frequency bands instead of the full frequency band as in MelGAN.
The predicted audio signals in each frequency band are upsampled and then passed to the synthesis filters.
These sum the signals from each frequency band to create the full-band audio signal.
Other improvements in MB-MelGAN include the expansion of the receptive field to about twice of that in MelGAN and substitute the feature matching loss with multi-resolution STFT loss, as in Parallel WaveGAN, which has been proven to be more effective to measure the difference between fake and real speech.

\subsection{Parallel WaveGAN}

Parallel WaveGAN is another parallel waveform generation method based on GAN \cite{pwgan}.
A WaveNet-based model conditioned on an auxiliary feature, e.g. mel-spectrogram, is used as the generator, which transforms input noise to the output waveform in parallel.
To improve the stability and efficiency of the adversarial training process, multi-resolution STFT loss is proposed. The STFT loss is sum of STFT losses with different analysis parameters, i.e. FFT size, window size and frame shift.
By combining multiple STFT losses with different analysis parameters, the generator can learn the time-frequency characteristics of speech better and it also prevents the generator from being overfit to a fixed STFT representation.

\subsection{MOS}

Mean Opinion Score (MOS) is the most commonly used method for assessing the quality of synthesised speech.
It is the arithmetic mean over all individual values on a predefined scale that a subject assigns to their opinion of the performance of a system.
It involves a group of subjects to listen to the different samples generated with one or several algorithms and evaluate them one by one on a scale.
The standard MOS scale ranges from 1 to 5, where 1 is lowest, and 5 is the highest perceived quality. 
Real human speech should score between 4.5 and 4.8.

The MOS score $\mu$ of a model is estimated by averaging the scores $m_k$ obtained for each $k$ of a set of synthesised samples $N$ using:
\begin{equation}
    \label{eq:mu}
\mu = \frac{1}{N} \sum \limits_{k=1}^{N} m_k \,\, .
\end{equation}
In addition, the 95\% confidence intervals for the scores are computed using:
\begin{equation}
\rm{CI} = \left[\mu - 1.96\frac{\sigma}{\sqrt{N}}, \mu + 1.96\frac{\sigma}{\sqrt{N}}\right] \,\, ,
\end{equation}
where $\sigma$ is the standard deviation of the scores.

\subsection{MUSHRA}

MOS can be readily used as a proxy to the perceptual quality of a TTS system. 
However, it is a time consuming process, making it hard to compare many models at the same time.
MUSHRA is an evaluation method used in codec listening test to evaluate lossy compression algorithms, and is defined by ITU-R recommendation BS.1534-3. 
In contrast to MOS, it makes it easier to compare different algorithms as all of the results are presented at the same time. 
In addition, since the scale is now in the range 0~--~100, more fine-grained differences can be captured.

In a MUSHRA listening test, the subject is presented with the reference, which is explicitly labeled so. 
Then, they are asked to rate a set of results that correspond to the reference.
The set also includes a hidden copy of the reference, as well as one or more anchors that should represent a lower or mid reference values.
The anchors serve the purpose of calibrating the score range so that small differences are not overly penalised.
These are typically a 3.5 kHz and a 7 kHz low-pass version of the reference. 

\section{Dataset}

To train the models, we first extracted a set of 3,500 phonetically rich utterances from a large body of Macedonian text comprising books, Wikipedia articles and online media.
We recorded a professional female speaker in the sound proof speech studio at FEEIT.
We used a studio-grade quality Sennheiser MK 4 condenser microphone to make the recordings.
The audio was sampled at 44.1 kHz with a 16 bit resolution using a professional Focusrite Scarlet 4i4 audio interface.
The total duration of the recorded speech corpus is 6 h.

\section{Experiment}

We trained the paramteric TTS model on the created speech corpus using the training tools provided by RHVoice.
We trained each the Deep Learning based models on the same data up to 250k iterations.

To evaluate their performance we synthesised a set of 15 phonetically rich utterances with varying length, using:
\begin{itemize}
    \item RHVoice,
    \item Tacotron 2 + Griffin-Lim,
    \item Tacotron 2 + MB-MelGAN, and
    \item Tacotron 2 + Parallel WaveGAN.
\end{itemize}
This resulted in a total of 60 synthesised utterances. 
In addition we also had our test utterances read by the recorded speaker. 

We then used the 10 shorter utterances for the MOS test, and the 10 longer one for the MUSHRA evaluation.
In this way we optimised the duration of the experiment, whilst keeping a 50\% overlap between the two tests.
This meant that listeners were asked to rate on the MOS scale 50 utterances, including the synthesised and natural recordings.
In the MUSHRA they were asked to compare 10 sets of stimuli, each set comprising of 6 stimuli including: the 4 synthesised utterances, the natural recording used as reference and a 3.5 kHz lowpass filtered anchor.

We invited 21 native speakers to part take in the listening experiments.
They were conducted online, using the webMUSHRA webapp \cite{webmushra}.
The listeners were instructed to wear headphones and be in a quiet room during the test.

\section{Results}

The results from the MOS analysis are presented in Table \ref{table_mos}.
We can see that the Parallel WaveGAN model scored slightly better than the RHVoice model.
On the other hand, the MB-MelGAN model scored the worst, even when compared to the Griffin-Lim based synthesis.
In fact, both of these models scored worse than the RHVoice model.

\begin{table}[bt]
\renewcommand{\arraystretch}{2.0}
\caption{Mean Opinion Score of the models}
\label{table_mos}
\centering
\begin{tabular}{lcc}
\hline
\bfseries \makecell{Model} & \bfseries \makecell{MOS} & \bfseries \makecell{95 \% CI}\\
\hline
RHVoice & 3.35 & 0.59 \\
Tacotron 2 + Griffin-Lim & 2.86& 0.67 \\
Tacotron 2 + MB-MelGAN & 2.20 & 0.52 \\
Tacotron 2 + Parallel WaveGAN & 3.43 & 0.58 \\
\hline
Natural speech & 4.79 & 0.30 \\
\hline
\end{tabular}
\end{table}

\begin{figure}[tbp]
\centerline{\includegraphics[width=\columnwidth]{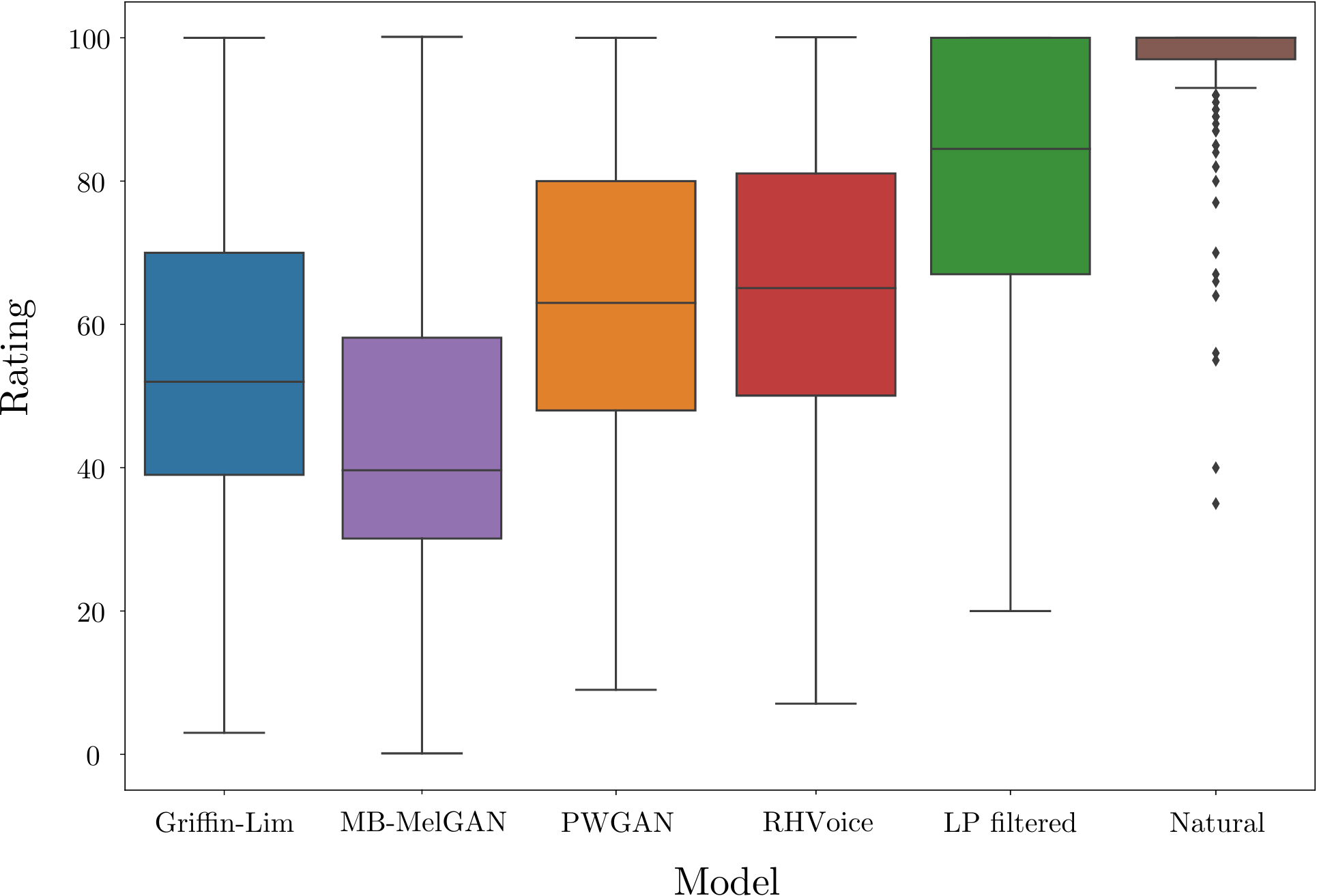}}
\caption{MUSHRA test results.}
\label{fig}
\end{figure}

The results from the MUSHRA analysis are shown in Fig.~\ref{fig}.
We can see that they confirm the results obtained with the MOS analysis, albeit RHVoice this time comes out on top by a small margin.
MB-MelGAN was evidently judged as the worse model.
What is interesting to note is that although the 3.5 kHz low-pass filtered natural speech was meant to serve as a lower bound anchor, in fact listeners preferred it to all of the models.
This was perhaps due to the markedly artificial sound of the synthesised samples, which was detrimental compared to a band-limited but natural signal.


\section{Conclusion}

In our work we trained and tested four speech synthesis models -- one based on paramteric speech synthesis and three based on Deep Learning.
The models were chosen so that they could be used for assistive devices and applications, including communication boards and screen readers.
We also designed and recorded a speech corpus with a professional speaker that was used for model training.
To evaluate the models we conducted both a MOS and a MUSHRA listening test.
The results show that the parametric HTS model fairs relatively well to the more advanced TTS models, scoring within a small margin with the more advanced Tacotron 2 + Parallel WaveGAN in both evaluations.
Since the parametric model requires a lot less resources, and it is also controllable both in speech rate and in pitch, we can conclude that it is the preferred choice for a Macedonian TTS system for the application scenario.

\nocite{fonetikaMANU}
\nocite{govorko}

\printbibliography

\end{document}